\documentclass{article}
\usepackage{spconf,amsmath,graphicx}
\usepackage{multirow,enumitem,adjustbox}
\usepackage{booktabs,threeparttable,makecell}
 \usepackage{amssymb}
 \usepackage[colorlinks=true,citecolor=green]{hyperref}
\usepackage{cleveref}

\usepackage{xcolor}
\usepackage{url}

\title{Disentangling the Prosody and Semantic Information with Pre-trained Model for In-Context Learning based Zero-Shot Voice Conversion}
\name{Zhengyang Chen$^1$, Shuai Wang$^{2,3}$, Mingyang Zhang$^{2,3}$, Xuechen Liu$^4$, Junichi Yamagishi$^4$, Yanmin Qian$^{1,\dagger}$
\thanks{$^\dagger$: Yanmin Qian is corresponding author}
}
\address{
$^1$Auditory Cognition and Computational Acoustics Lab \\
MoE Key Lab of Artificial Intelligence, AI Institute \\
Department of Computer Science and Engineering, Shanghai Jiao Tong University, Shanghai China \\
$^2$Shenzhen Research Institute of Big Data, Shenzhen, China \\
$^3$Chinese University of Hong Kong, Shenzhen, China \\
$^4$National Institute of Informatics, Tokyo, Japan \\
}

\begin{document}
\ninept
\maketitle
\begin{abstract}
Voice conversion (VC) aims to modify the speaker's timbre while retaining speech content. Previous approaches have tokenized the outputs from self-supervised into semantic tokens, facilitating disentanglement of speech content information. Recently, in-context learning (ICL) has emerged in text-to-speech (TTS) systems for effectively modeling specific characteristics such as timbre through context conditioning. This paper proposes an ICL capability enhanced VC system (ICL-VC) employing a mask and reconstruction training strategy based on flow-matching generative models. Augmented with semantic tokens, our experiments on the LibriTTS dataset demonstrate that ICL-VC improves speaker similarity. Additionally, we find that k-means is a versatile tokenization method applicable to various pre-trained models. However, the ICL-VC system faces challenges in preserving the prosody of the source speech. To mitigate this issue, we propose incorporating prosody embeddings extracted from a pre-trained emotion recognition model into our system. Integration of prosody embeddings notably enhances the system's capability to preserve source speech prosody, as validated on the Emotional Speech Database.


\end{abstract}
\section{Introduction}
\label{sec:intro}

Human speech contains various aspects of information, such as content, speaker timbre, and speaking style.
Voice conversion (VC) allows the alteration of one's voice to mimic another's characteristics without changing the spoken content and speaking style~\cite{liu20v_interspeech,wang2023delivering}. This technology finds diverse applications across various sectors, such as dubbing for movies, timbre transformation in online live streaming, and identity anonymization.

To achieve effective voice conversion, it is crucial to separate the content information from the source speech. Pre-trained automatic speech recognition (ASR) models~\cite{sun2016phonetic} and self-supervised pre-trained models~\cite{choi2021neural,lin2021fragmentvc,DBLP:conf/iclr/YinRLWXZ22,polyak21_interspeech} are typically utilized to extract linguistic content. To minimize the non-content information, outputs from self-supervised pre-trained models are often tokenized to generate semantic tokens~\cite{zhang2024speechlm,zhou2024phonetic,yang2024towards,chen2024loss,gong2023zmm,wang2024ham}. However, tokenization methods are usually tied to the training objectives of the self-supervised models, resulting in differing token formats across different pre-trained models. This necessitates the design of distinct modules to accommodate various semantic token formats. Interestingly, despite the differing training objectives of these models, we discovered that the k-means method can serve as a universal tokenization strategy. This approach standardizes the semantic token format across different pre-trained models.

Voice conversion technology has significantly evolved over the past few decades, advancing from traditional methods to cutting-edge zero-shot voice conversion. Zero-shot voice conversion represents a paradigm shift, allowing models to convert a source voice to a target voice they have never encountered during training. By leveraging advancements in speaker embedding techniques, zero-shot voice conversion systems can generalize to new, unseen speakers without requiring fine-tuning with additional data~\cite{qian2019autovc}.

However, utilizing speaker embeddings to represent timbre information in zero-shot voice conversion demands a robust pre-trained speaker encoder~\cite{cooper2020zero}. Additionally, specific modules must be designed to incorporate the information from speaker embeddings into the VC system. Both text-to-speech (TTS) and voice conversion training typically require high-fidelity audio data~\cite{zen2019libritts,koizumi2023libritts}, which limits the scalability of the system to large-scale web data. Recently, researchers have proposed leveraging in-context learning (ICL) capability ~\cite{wang2023neural,du2024unicats,le2024voicebox,anastassiou2024seed} to enable TTS systems to synthesize speech for unseen speakers through a target speech prompting strategy. This approach eliminates the need for a pre-trained speaker encoder and has no stringent audio quality requirements.

\begin{figure*}[ht!]
    \centering
    \includegraphics[width=.95\textwidth]{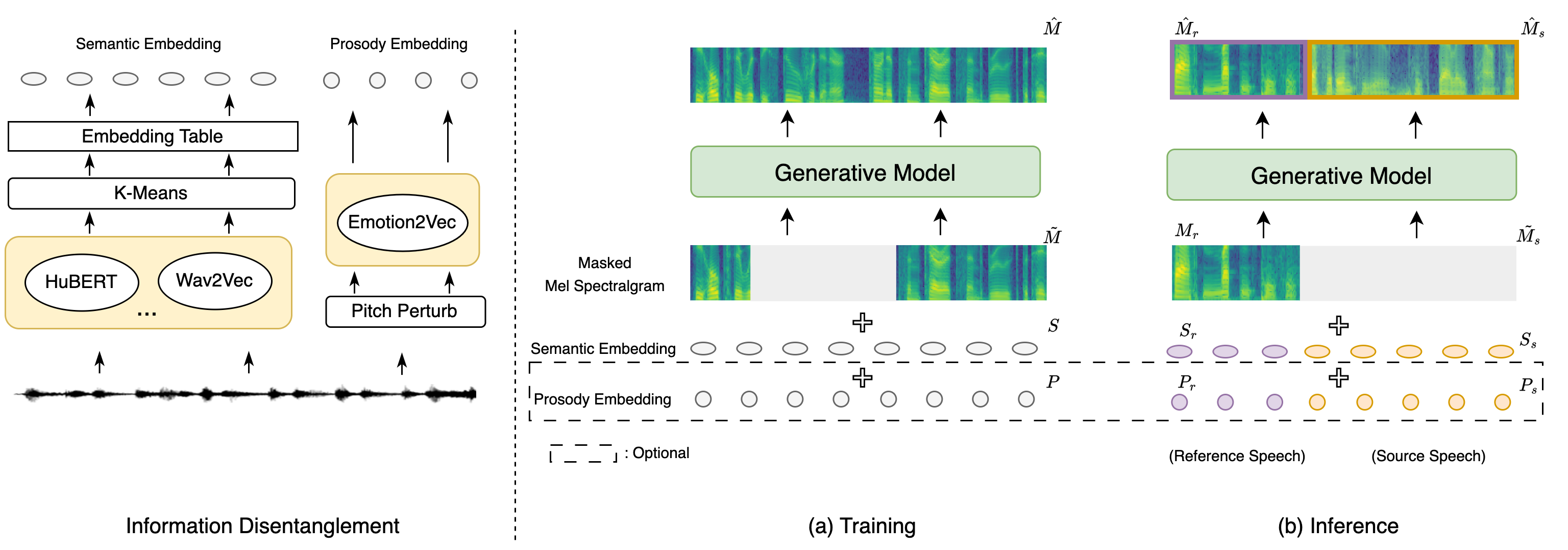}
    \caption{System Overview. 
    }
    \label{fig:system_overview}
\end{figure*}

In this paper, we equip the voice conversion system with the ICL capability to propose ICL-VC. Initially, we disentangle the content information from speech by extracting semantic tokens using a self-supervised pre-trained model. Next, we train the model through a mel-spectrogram mask and reconstruction strategy based on flow-matching generative model to equip the model with the ICL capability. During inference, we perform voice conversion by prompting the system with reference speaker speech. This strategy allows the system to achieve zero-shot voice conversion with high speaker similarity. However, leveraging ICL requires the VC system to extract both timbre and prosody information from the reference speech, which can compromise the preservation of the source speech's prosody. To address this issue, a straightforward solution is to incorporate prosody information, such as pitch and energy, into the VC system. However, we found that even normalized pitch and energy can lead to timbre information leakage and limit the system's ability to handle prosody-unmatched scenarios. To enhance the prosody preservation capability of the ICL-VC system, we propose disentangling prosody information using the Emotion2Vec~\cite{ma2023emotion2vec} pre-trained emotion recognition model, which significantly improves system performance. Besides, we also provide some samples at \url{https://czy97.github.io/ICL-VC/}.

\section{Method}

\subsection{In-context Learning based Voice Conversion}
\label{ssec:system_overview}

An overview of our system is shown in Figure \ref{fig:system_overview}. For the voice conversion task, it is crucial to preserve the content of the source speech. Given an audio segment $\mathbf{x} = [x_1, x_2, \ldots, x_T] \in \mathbb{R}^T$, we first disentangle the content information by extracting the semantic embedding sequence $S = [\mathbf{s}_1, \mathbf{s}_2, \ldots, \mathbf{s}_{T'}] \in \mathbb{R}^{T' \times d}$, as described in Section \ref{ssec:semantic}. Additionally, we extract the mel-spectrogram $M = [\mathbf{m}_1, \mathbf{m}_2, \ldots, \mathbf{m}_{T'}] \in \mathbb{R}^{T' \times f}$ from the audio segment $\mathbf{x}$. During training, we employ a mask and reconstruction paradigm. As illustrated in Figure \ref{fig:system_overview}, we mask part of the mel-spectrogram $M$ with random noise to obtain $\tilde{M}$. The semantic embedding sequence $S$ is then concatenated with the masked mel-spectrogram $\tilde{M}$ to form the model input $(S; \tilde{M}) \in \mathbb{R}^{T' \times (d+f)}$. A generative model $G$ is used to reconstruct the original mel-spectrogram: $M \approx \hat{M} = G(S; \tilde{M})$. In this learning paradigm, the generative model must learn to extract content information from the semantic embedding and contextual information from the unmasked parts of the mel-spectrogram to reconstruct the masked sections, demonstrating in-context learning (ICL) capabilities.

In the inference stage, we denote the semantic embeddings extracted from the reference and source speech as $S_r \in \mathbb{R}^{T_r' \times d}$ and $S_s \in \mathbb{R}^{T_s' \times d}$, respectively, and concatenate them along the time dimension to form $S_{\text{infer}} \in \mathbb{R}^{(T_r' + T_s') \times d}$. The mel-spectrogram extracted from the reference speech is denoted as $M_r \in \mathbb{R}^{T_r' \times f}$, and we mask all parts of the mel-spectrogram extracted from the source speech with random noise to obtain $\tilde{M}_s \in \mathbb{R}^{T_s' \times f}$. The model $G$ then generates the new mel-spectrogram $\hat{M}_s$ for the corresponding source speech input: $[\hat{M}_r, \hat{M}_s] = G(S_{infer}; [M_r, \tilde{M}_s])$. When generating the mel-spectrogram $\hat{M}_s$, the model conditions on the semantic embedding $S_s$ from the source speech and extracts speaker timbre information from the reference mel-spectrogram $M_r$ using its ICL capability. This approach ensures that the generated mel-spectrogram contains the content of the source speech but the timbre of the reference speech. Thus, voice conversion is achieved, and we name this strategy in-context learning based voice conversion (ICL-VC).

In some application scenarios, such as voice conversion for video, it is crucial to preserve the prosody information to match the human lip movements and facial expressions. However, using the ICL capability described above, the system infers the prosody information for the generated mel-spectrogram from the reference speech's mel-spectrogram. This may result in generated speech with prosody different from the source speech, which is also verified by our experiment results in Table \ref{table:esd_res}. To maintain the prosody of the source speech, we propose adding the prosody embedding $P = [\mathbf{p}_1, \mathbf{p}_2, \ldots, \mathbf{p}_{T'}] \in \mathbb{R}^{T' \times h}$, as introduced in Section \ref{ssec:prosody_intro}, to the system's input. This enables the system to directly extract prosody information from the source speech's prosody embedding during inference to generate the mel-spectrogram.


\subsection{Semantic Information Extraction}
\label{ssec:semantic}

To extract the semantic embedding, we utilize self-supervised learning-based pre-trained models such as HuBERT~\cite{hsu2021hubert} and Wav2Vec~\cite{baevski2020wav2vec}. To remove as much non-content information as possible, we tokenize the output from the pre-trained model to extract semantic tokens and then use a learnable embedding module to map the tokens to embeddings. It should be noted that prior research~\cite{zhang2024speechlm,zhou2024phonetic,yang2024towards,chen2024loss,gong2023zmm} has employed different pre-trained models for various application scenarios. HuBERT is frequently used due to its straightforward k-means tokenization method. Additionally, researchers prefer Wav2Vec for multilingual scenarios~\cite{gong2023zmm} because of the availability of the Wav2Vec multilingual pre-trained model\footnote{The Wav2Vec XLSR-53 model has been trained on 53 languages.}\footnote{\url{https://github.com/facebookresearch/fairseq/tree/main/examples/wav2vec}}. However, the original tokenization strategy in Wav2Vec generates two semantic token sequences~\cite{baevski2020wav2vec,gong2023zmm}, which is more complex and differs from the single semantic token sequence produced by HuBERT. This difference in semantic tokens from various pre-trained models complicates the adaptation of a single system to different semantic tokens without modifying the model architecture. The reason researchers choose different tokenization methods for different pre-trained models is to match the pre-trained models' training objectives~\cite{hsu2021hubert,baevski2020wav2vec}. 
Interestingly, we found that k-means is a versatile tokenization method applicable to different self-supervised pre-trained models regardless of their training objectives. In our experiments, we propose leveraging the k-means method for semantic tokenization across various pre-trained models.



\subsection{Prosody Information Extraction}
\label{ssec:prosody_intro}


As introduced in Section \ref{ssec:system_overview}, for certain application scenarios, we need to provide the system with additional prosody embedding to maintain the source speech's prosody. Prosody in speech is primarily reflected in the pitch, energy, and duration of each phoneme's articulation~\cite{du22b_interspeech}. In our ICL-VC framework, we can directly ensure that the duration of the generated speech matches the source speech. Therefore, the aspects of prosody that need to be considered are mainly pitch and energy. Previous work~\cite{qian2020f0,du22b_interspeech,liu23d_interspeech,mitsui2023towards} directly leveraged normalized pitch (F0) or energy information to guide speech synthesis. However, in our experiments, we found that even when F0 and energy values are normalized using mean and standard deviation, they may still cause speaker information leakage and do not generalize well to prosody-unmatched voice conversion tasks.

To mitigate this problem, we propose extracting prosody embeddings from the pre-trained Emotion2Vec~\cite{ma2023emotion2vec} model. The Emotion2Vec model is trained to capture emotion information in speech, which we believe is primarily reflected in speech prosody. Thus, we hypothesize that the representations extracted from the Emotion2Vec model can serve as prosody embeddings in our system. Additionally, to avoid speaker timbre information leakage from the prosody embedding, we first perturb the pitch~\cite{yamamoto2019speaker} of the speech before feeding it into the Emotion2Vec model using the sox\footnote{\url{https://linux.die.net/man/1/sox}} toolkit to raise or lower the pitch by 200 or 400 cents. Then, we use the final output from the Emotion2Vec model as the prosody embedding.


\subsection{Flow-Matching based Generative Model}

To balance the quality and speed of mel-spectrogram generation, we employ the flow-matching-based~\cite{DBLP:conf/iclr/LipmanCBNL23} generation method in our study. This method aims to fit an unknown distribution $q(x_1)$, where $x_1 \in \mathbb{R}^d$, by constructing a continuous flow $\phi_t: \mathbb{R}^d \rightarrow \mathbb{R}^d, t \in [0, 1]$ that transforms a simple known distribution $p_0$, such as a Gaussian distribution, into the target distribution $p_1 \approx q$. The flow is governed by an ordinary differential equation (ODE):
\begin{equation}
    \frac{d}{dt} \phi_t(x) = u_t\left(\phi_t(x)\right)
\end{equation}

To construct the flow model, we can approximate the vector field $u_t \in \mathbb{R}^d$ within the ODE equation. However, a closed-form formulation for $u_t$ does not exist. Previous work~\cite{DBLP:conf/iclr/LipmanCBNL23} has shown that approximating a conditional vector field $u_t(x|x_1)$ is equivalent to approximating the original $u_t$, leading to the development of a conditional flow matching (CFM) training objective:
\begin{equation}
\label{eq:CFM}
\mathcal{L}_{\mathrm{CFM}}(\theta) = \mathbb{E}_{t, q(x_1), p_t(x|x_1)} \left\| v_t(x, \theta) - u_t(x|x_1) \right\|^2
\end{equation}
where $p_t(x|x_1)$ denotes the conditional probability density function at time $t$, and $v_t(x, \theta)$ is a time-dependent neural network used to approximate $u_t(x|x_1)$. In the training process, $t$ is randomly sampled from $[0,1]$ and we encode it as a sinusoidal positional embedding, which is concatenated with the model input. Additionally, we adopt the optimal transport (OT) path introduced in~\cite{DBLP:conf/iclr/LipmanCBNL23} to define the flow, where $p_t(x|x_1) = \mathcal{N}(x | t x_1, (1 - (1 - \sigma_{\min}) t)^2 I)$ and $u_t(x|x_1) = (x_1 - (1 - \sigma_{\min}) x) / (1 - (1 - \sigma_{\min}) t)$. Here, $\sigma_{\min}$ is a small scalar marginally above zero and is set to $10^{-5}$ in our experiment.





To implement the flow-matching-based generative model in our system, we extend $v_t$ to condition on the mel-spectrogram generated at each timestamp $M_t$, semantic embedding $S$, and optionally the prosody embedding $P$, resulting in $v_t(M_t, S, P; \theta)$. During training, we only focuses on optimizing the masked part:

\begin{equation}
\label{eq:icl_vc}
\begin{aligned}
\mathcal{L}(\theta) &= \mathbb{E}_{t, q(M), p_t(M_t|M)} \Big\Vert m(\Omega_t) \Big\Vert^2 \\
\Omega_t &= v_t(M_t, S, P; \theta) - u_t(M_t|M)
\end{aligned}
\end{equation}
where $m$ is a mask function that will mask the unmasked part of input mel-spectrogram $\tilde{M}$.

In the inference stage, we integrate the following ODE equation \ref{eq:infer_ode} from $t=0$ to $t=1$ to obtain the predicted mel-spectrograms $\hat{M}_r$ and $\hat{M}_s$, with only $\hat{M}_s$ retained for the voice conversion task:
\begin{equation}
\label{eq:infer_ode}
\begin{aligned}
\frac{d}{dt} \phi_t(M_t) &= v_t(M_t, S_{\text{infer}}, P_{\text{infer}}; \theta) \\
\phi_0 &= [M_r, \tilde{M}_s]
\end{aligned}
\end{equation}
To achieve a balance between generative fidelity and time consumption, we configured the ODE step to 32 in our experiment.

    


\section{Experimental Setup}

\subsection{Dataset}
In our experiment, we utilized two datasets: the LibriTTS~\cite{zen2019libritts} dataset and the Emotional Speech Database (ESD)~\cite{zhou2022emotional}. The train-clean-360 subset of the LibriTTS dataset was used for system training, while the test-clean subset was employed to evaluate the basic zero-shot voice conversion ability of our ICL-VC system. The ESD dataset, which contains speech with rich prosody variations, was exclusively used to assess the source speech prosody-preserving ability. Additionally, we resampled the ESD dataset to 24 kHz to match the sample rate of the LibriTTS dataset.

For evaluating the basic voice conversion ability, we randomly selected 20 speakers from the test-clean subset of the LibriTTS dataset. Ten speakers were designated as reference speakers and another ten as source speakers. For each reference speaker, we randomly selected one utterance, and for each source speaker, we randomly selected two utterances. We then converted the timbre of each source speaker's utterance to match all the reference speakers, resulting in a total of 200 generated utterances for testing.

To evaluate the prosody-preserving ability of the VC systems, we randomly selected 16 utterances from the English portion of the ESD dataset as the source speech. The ten reference utterances selected in the previous evaluation were also used in this assessment.

\subsection{Model Configuration}

We implemented our generative model as an encoder-only Transformer model~\cite{vaswani2017attention}. The Transformer consists of 8 layers, with an input dimension set to 768. A linear layer is inserted before the Transformer to adjust the input feature dimension to match the Transformer's input dimension. Additionally, another linear layer maps the Transformer's output to the dimension of the mel-spectrogram. We extract the mel-spectrogram using the Vocos~\cite{siuzdak2023vocos} toolkit and employ its pre-trained mel-based\footnote{\url{https://github.com/gemelo-ai/vocos}} vocoder to convert the mel-spectrogram into raw audio in the inference stage.

To extract prosody information, we use the Pyworld~\footnote{\url{https://pypi.org/project/pyworld/}} toolkit to extract the pitch (F0) from the audio. We normalize the pitch and energy values for each utterance to mitigate speaker information leakage. The pitch and energy values are then tokenized into 256 bins, producing pitch and energy tokens. A learnable embedding module maps these tokens into embeddings. When using the Emotion2Vec~\cite{ma2023emotion2vec} model to extract prosody information, as introduced in Section \ref{ssec:prosody_intro}, we first perturb the pitch of the input speech and use the final output of the Emotion2Vec model as the prosody embedding.

In our experiment, we utilize three pre-trained models to extract semantic tokens following the fairseq\footnote{\url{https://github.com/facebookresearch/fairseq/tree/main/examples/hubert/simple_kmeans}} recipe: the HuBERT base model trained on LibriSpeech-960, the Wav2Vec base model trained on LibriSpeech-960, and the Wav2Vec-XLSR model trained on 56k hours of multilingual data. For each pre-trained model, we train a k-means tokenizer with 500 clusters separately. Previous research~\cite{chen2022wavlm,chen2022large} has demonstrated that different layers of self-supervised pre-trained models contain varying types of information. Based on this analysis, we select outputs from the 9th layer of both the HuBERT base and Wav2Vec base models, and the 14th layer of the Wav2Vec-XLSR model, to train the k-means tokenizer. This layer selection aims to retain the most semantic information while filtering out non-semantic information.

During the training process of our ICL-VC system, we randomly select an unmasked region of each input utterance, with a duration ranging from 2 to 3 seconds. The left part of this region is then masked with random noise. The system is trained to reconstruct the masked portion by conditioning on the input semantic embedding (or with addition prosody embedding) and the unmasked region.

\subsection{Baseline System}
For system comparison, we introduce two baseline systems in our experiment. The first one is the YourTTS system~\cite{casanova2022yourtts}, where we use the ResNet34-based pre-trained speaker encoder provided by Wespeaker toolkit~\cite{wang2023wespeaker}. The other one is RefXVC~\cite{zhang2024refxvc}, a voice conversion method leveraging self-supervised learning features and enhanced reference-based speaker embedding to improve speaker similarity. 
All the systems in our experiment are trained on the same training set for fair comparison.

\subsection{Evaluation Metric}
\label{ssec:evaluation_metric}

In our experiments, we conduct both objective and subjective evaluations. For the objective evaluation, we compute the speaker embedding cosine similarity (SECS) between the converted speech and the reference speech. We use the ECAPA-TDNN~\cite{desplanques2020ecapa} pre-trained model from the Wespeaker toolkit~\cite{wang2023wespeaker} to extract the speaker embeddings. Additionally, following~\cite{hussain2023ace}, we utilize a ASR pre-trained model provided by the NEMO toolkit\footnote{\url{https://catalog.ngc.nvidia.com/orgs/nvidia/teams/nemo/models/stt_en_quartznet15x5}} to evaluate the character error rate (CER) of the ground-truth and converted speech, which reflects the intelligibility of the speech. Furthermore, we calculate the Pearson correlation between the pitch and energy sequences~\cite{ning2023expressive} extracted from the source speech and the converted speech to determine if the converted speech maintains the prosody of the source speech. For the subjective evaluation, we ask 15 human raters to provide a mean opinion score (MOS) ranging from 1 to 5 regarding the naturalness of the speech and the prosody similarity between the converted and source speech. These 15 human raters simultaneously participated in scoring both naturalness MOS and prosody MOS. When conducting a prosody similarity subjective evaluation, the instruction we provided is: ``Please assess the similarity in speaking prosody between the provided speech and the reference speech. Give a rating from 1 to 5, with 5 indicating completely consistent prosody and 1 indicating completely different prosody." On average, each audio sample receives 3 MOS scores for naturalness evaluation and 4 MOS scores for prosody similarity evaluation.

\section{Results}
\subsection{Zero-shot Voice Conversion Evaluation}

\subsubsection{Comparison between Different Systems}
In this section, we evaluate our proposed ICL voice conversion strategy on the LibriTTS test set, with the results summarized in Table \ref{table:libritts_res}. Our findings indicate that all ICL based VC systems utilizing different semantic codecs demonstrate superior performance in replicating the target speaker's timbre and generating a more natural voice. As introduced in Section \ref{ssec:semantic}, unlike other works~\cite{gong2023zmm} that use the original tokenization strategy in Wav2vec, we employ a simple k-means tokenization strategy for Wav2vec. The results reveal that the k-means tokenization strategy is effective for the Wav2Vec framework, despite Wav2Vec not being trained with labels clustered by the k-means method. This phenomenon suggests that the k-means tokenization method could be a general strategy for semantic tokenization.

Interestingly, the ICL-VC-Wav2Vec-XLSR system exhibits better cosine speaker similarity compared to ICL-VC-HuBERT and ICL-VC-Wav2Vec. This may be attributed to the Wav2Vec-XLSR model being trained on a larger amount of speech data and exposed to more speakers, thus enhancing its zero-shot voice conversion capability. However, the ICL-VC-Wav2Vec-XLSR system shows a higher Character Error Rate (CER) than the other two systems. This discrepancy is likely because the HuBERT and Wav2Vec models are directly trained on the LibriSpeech dataset, which has a high content correlation with the LibriTTS test set.

\begin{table}[ht!]
\centering
\caption{\textbf{Voice Conversion Results on LibriTTS dataset.} Prosody embedding is not used for ICL-VC system's input in this evaluation.
}
\begin{adjustbox}{width=.48\textwidth,center}
\begin{threeparttable}
\begin{tabular}{l|ccc}
\toprule
System & SECS & CER & Natualness MOS \\
\hline
YourTTS & 0.824 & 3.83 & 3.86 $\pm$ 0.13\\
RefXVC & 0.797 & 2.22 & 4.02 $\pm$ 0.12 \\
\hline
ICL-VC-HuBERT & 0.846 & 2.66 &  4.33 $\pm$ 0.11\\
ICL-VC-Wav2Vec & 0.845 & 2.23 & 4.37 $\pm$ 0.12 \\
ICL-VC-Wav2Vec-XLSR & 0.856 &  5.09 & 4.32 $\pm$ 0.11 \\
\hline
Ground Truth & - & 1.69 & 4.51 $\pm$ 0.09 \\

\bottomrule
\end{tabular}
\end{threeparttable}
\label{table:libritts_res}
\end{adjustbox}
\end{table}

\begin{table*}[ht!]
\centering
\caption{\textbf{Voice Conversion Results on ESD dataset.} In this experiment, only the semantic tokens extracted from HuBERT pre-trained model are used in our ICL-VC system. The Pitch Corr and Energy Corr correspond to the Pearson correlation introduced in section \ref{ssec:evaluation_metric}.
}
\begin{adjustbox}{width=.8\textwidth,center}
\begin{threeparttable}
\begin{tabular}{l|cccccc}
\toprule
System & SECS & CER &  Pitch Corr & Energy Corr & Prosody MOS & Naturalness MOS \\
\hline
YourTTS & 0.706 & 10.68 &  0.683 & 0.898 & 4.02 $\pm$ 0.22 & 3.33 $\pm$ 0.15 \\
RefXVC & 0.748 & 16.50 &   0.512 & 0.695 & 3.79 $\pm$ 0.13 & 3.53 $\pm$ 0.12 \\
\hline
ICL-VC & 0.800 & 5.39  & 0.608 & 0.867 & 3.34 $\pm$ 0.18 & 3.77 $\pm$ 0.17 \\
\hspace{3pt} + Pitch \& Energy & 0.769 & 5.31 &   0.727 & 0.935 & 4.12 $\pm$ 0.17 & 3.61 $\pm$ 0.16\\
\hspace{3pt} + Emotion Emb & 0.789 & 4.33 &  0.671 &   0.900 & 4.19 $\pm$ 0.16 & 4.14 $\pm$ 0.13\\
\hline
Ground Truth & - & 3.21 &  - & - & - & 4.41 $\pm$ 0.08 \\

\bottomrule
\end{tabular}
\end{threeparttable}
\label{table:esd_res}
\end{adjustbox}
\end{table*}

\subsubsection{Comparison between Different Reference Durations}
\begin{figure}[ht!]
    \centering
    \includegraphics[width=\linewidth]{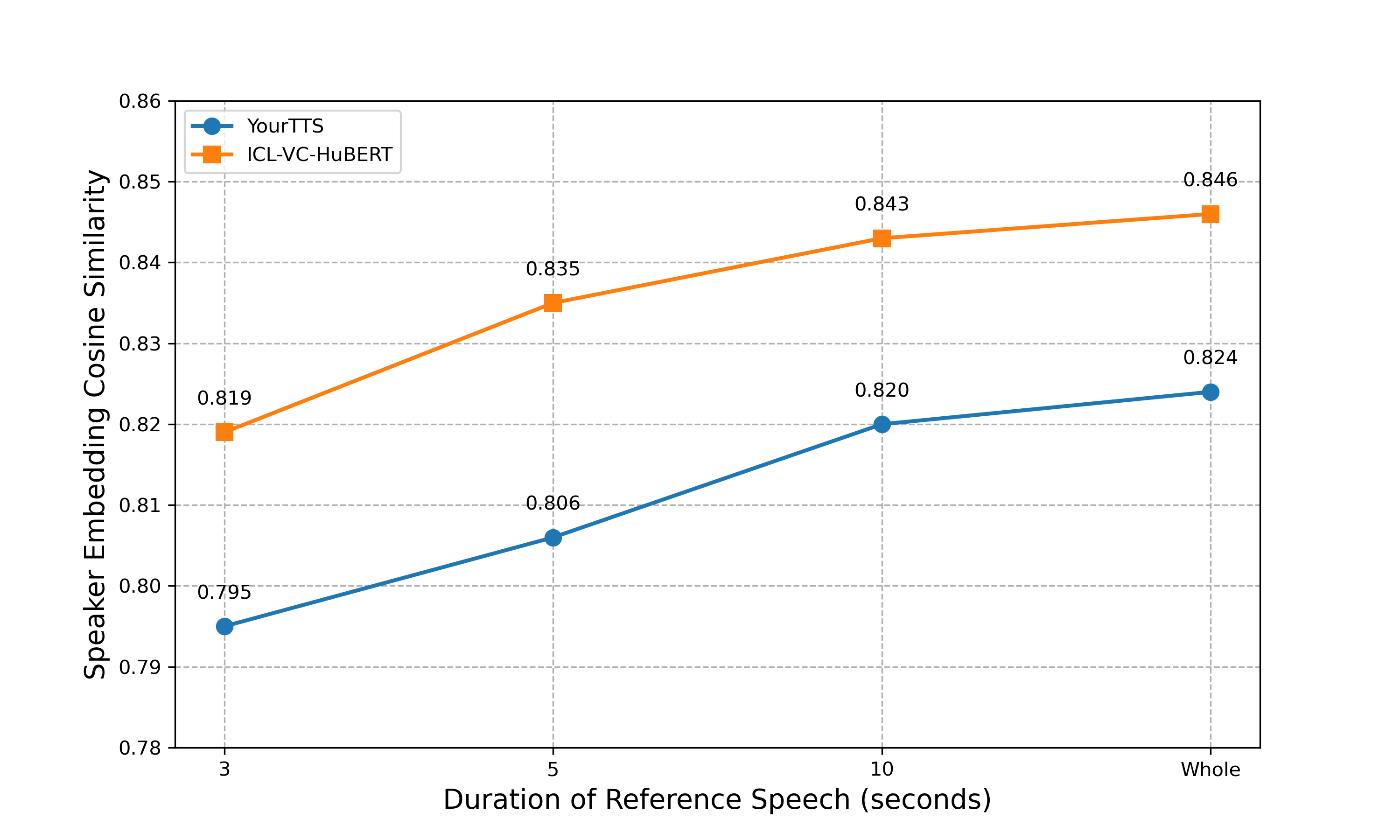}
    \caption{The relationship between the speaker embedding cosine similarity (SECS) and reference speech duration.
    }
    \label{fig:duration_and_similarity}
\end{figure}

Evaluating the use of shorter reference speech for timbre replication is essential to testing the capabilities of zero-shot voice conversion systems. In this section, we evaluate the impact of reference duration on speaker embedding cosine similarity, as illustrated in Figure \ref{fig:duration_and_similarity}. For both YourTTS and our method, the SECS variation trend indicates that longer reference speech allows the system to better model the speaker's timbre information. Notably, our ICL-VC system achieves high cosine speaker similarity even with a very short (3-second) reference utterance.

\subsection{Prosody Preserving Ability Evaluation}
In addition to mimicking the target speaker's timbre, the voice conversion task always requires preserving the source speech's content and prosody~\cite{liu20v_interspeech}. To evaluate the system's ability to maintain prosody, we directly evaluate the system trained on LibriTTS on the ESD dataset, which features speech with significant prosody variations. Beyond the evaluation metrics listed in Table \ref{table:libritts_res}, we incorporate pitch and energy correlation metrics introduced in section \ref{ssec:evaluation_metric} to assess the prosody similarity between the converted and source speech. Additionally, we include prosody MOS scores to capture subjective human judgments of prosody similarity. The corresponding results are shown in Table \ref{table:esd_res}. It is noteworthy that all SECS values in Table \ref{table:esd_res} are lower than those in Table \ref{table:libritts_res}, due to the prosody mismatch between LibriTTS and ESD speech. Nevertheless, these SECS values remain in the high range, indicating that all systems are capable of performing voice conversion on the ESD dataset.

Despite the superior timbre replication capability of the ICL-based VC system, the results in Table \ref{table:esd_res} indicate that it struggles to maintain prosody information. The ICL-VC system exhibits lower prosody MOS scores compared to the two baseline systems, and lower pitch and energy correlation than the YourTTS system. This shortcoming arises because, during inference, our ICL-VC system relies on the reference speech to extract both timbre and prosody information, while the source speech provides only semantic information. To enhance our ICL-VC system's ability to retain prosody, we employ two strategies introduced in Section \ref{ssec:prosody_intro}. First, we integrate normalized pitch and energy information into the input of our ICL-VC system. This approach significantly improves prosody correlation and MOS scores, although it leads to a notable decline in speaker similarity and naturalness MOS scores. While normalized pitch reduces timbre leakage from the source speech to the converted speech, the ESD dataset's prosody variations differ greatly from those in the LibriTTS dataset. This discrepancy can cause mismatches during inference, impairing the model's ability to generate high-quality speech and resulting in lower SECS and naturalness MOS scores. Interestingly, our proposed strategy using prosody information extracted from the pre-trained Emotion2Vec model helps mitigate this issue, yielding higher SECS and naturalness MOS scores.

\section{Conclusion}
In this paper, we introduced the in-context learning (ICL) method to the voice conversion task. Our experimental results demonstrate that the ICL method enables the system to perform voice conversion without relying on an external speaker encoder, resulting in higher speaker similarity. However, we observed that simply applying the ICL method does not ensure the preservation of the source speech's prosody. To address this issue, we proposed extracting prosody embeddings from a pre-trained emotion recognition model, achieving impressive prosody-preserving performance on the ESD dataset. Additionally, we found that the k-means tokenization method is a versatile technique that can be applied to various self-supervised pre-trained models, regardless of their different training objectives.
Despite these advancements, our method has been validated only on small datasets so far. The ICL method, combined with the information decoupling approach based on self-supervised learning pre-trained models, has the potential to train the system on larger, noisier, stylistically diverse, and even multilingual datasets. This will be the focus of our future work.

\section{ACKNOWLEDGMENTS}
\label{sec:ack}
This work was supported in part by China NSFC projects under Grants 62122050 and 62071288, in part by Shanghai Municipal Science and Technology Commission Project under Grant 2021SHZDZX0102. This study is partially supported by MEXT KAKENHI Grants (24K21324) and CCF-NetEase ThunderFire Innovation Research Funding (CCF-Netease 202302).


\newpage

\bibliographystyle{IEEEbib}
\bibliography{refs}

\end{document}